
----------
X-Sun-Data-Type: default
X-Sun-Data-Description: default
X-Sun-Data-Name: preprint2.tex
X-Sun-Content-Lines: 544


\baselineskip=13pt
\magnification=\magstep1
\def\gtorder{\mathrel{\raise.3ex\hbox{$>$}\mkern-14mu
             \lower0.6ex\hbox{$\sim$}}}
\def\ltorder{\mathrel{\raise.3ex\hbox{$<$}\mkern-14mu
             \lower0.6ex\hbox{$\sim$}}}
\def\proptwid{\mathrel{\raise.3ex\hbox{$\propto$}\mkern-14mu
             \lower0.6ex\hbox{$\sim$}}}

\font\maintitle= cmbx10 at 15.00truept
\font\sechead= cmbx10 at 12.00truept
\bigskip
\centerline{\maintitle GAMMA-RAY BURSTS FROM}
\centerline{\maintitle BLAST WAVES AROUND GALACTIC NEUTRON STARS}

\bigskip
\medskip
\centerline{Mitchell C. Begelman\footnote{$^1$}{Joint Institute for Laboratory
Astrophysics,
University of Colorado and}\footnote{}{National Institute of Standards and
Technology, Boulder,
CO 80309}$^,$\footnote{$^2$}{Department of Astrophysical,
Planetary and Atmospheric Sciences, Uni-}\footnote{}{versity of Colorado,
Boulder, CO
80309.}, P. M\'esz\'aros\footnote{$^3$}{Pennsylvania State University, 525
Davey Lab,
University Park, PA 16803.}, and Martin J. Rees\footnote{$^4$}{Institute of
Astronomy,
Madingley Road, Cambridge CB3 0HA, England.} }

\bigskip
\centerline{\it Submitted to M.N.R.A.S. Pink Pages, \underbar{Aug. 4, 1993} ;
in press. }
\bigskip

\centerline{ABSTRACT}

\medskip

\bigskip

\centerline{\sechead 1.  INTRODUCTION}

\medskip

The remarkable isotropy of classical gamma-ray bursts (GRBs) sampled by the
BATSE
instrument on GRO (Meegan et al.~1992; Fishman et al.~1993) requires their
sources to be
located at cosmological distances, in an extended halo of the Galaxy, or
in a more local distribution ($\ltorder 1-2$ Kpc) which contrives to
be almost isotropic about us.
Many authors have attributed "galactic" bursts to
violent disturbances in the magnetospheres of neutron stars (e.g., Blaes et
al.~1990; Ramaty et
al.~1981); however, the details of the emission  mechanisms have generally been
left
unspecified.  While the tenuous plasma filling the magnetosphere may be
responsible for much
of the high-energy radiation, any  disturbance of the kind proposed is also
likely to expel
magnetic flux and plasma into the interstellar medium (ISM) surrounding the
neutron star,
possibly at relativistic speed.  In this Note, we point out that the blast wave
driven into the ISM
by a magnetospheric disturbance could also produce a flash of gamma-rays with
the
characteristics observed to be typical of GRBs.  We tentatively associate these
two modes of
emission with the short/variable and long/smooth subgroups of bursts,
respectively, which have
recently been identified through analyses of BATSE data (Kouveliotou et
al.~1993; Lamb,
Graziani, \& Smith 1993). Moreover, bursts associated with emission from blast
waves
would become more conspicuous in the gamma-ray band when they occur in a denser
environment, thus providing a possible explanation for modulation associated
with spiral
arm structure, as has been claimed by Quashnock \& Lamb (1993).

\bigskip

\centerline{\sechead 2.  INTERSTELLAR ENVIRONMENT OF AN ISOLATED NEUTRON
STAR}

\medskip

Although neutron stars are believed to begin their lives in the rarefied
stellar wind
bubbles and supernova remnants created by their progenitor stars, within about
$10^5$
yr they will be interacting with the general ISM (Shull, Fesen, \& Saken 1989).
 Even for
modest values of the surface dipole field $10^{11} B_{11}$ G and spin period
$P$ s,
electromagnetic forces will prevent the ambient interstellar gas (of density
$n_\infty$) from
reaching the surface of the neutron star.  When gravitational focusing is
unimportant, the dipole
spindown energy loss will create a standoff bow shock in the ISM with a radius
$$ r_W \sim 8\times 10^{13}  { B_{11} \over P^2 v_{100} n_\infty^{1/2} } \
{\rm cm} ,
\eqno(1) $$
where $100 v_{100}$ km s$^{-1}$ is the (supersonic) speed of the neutron star
through
the ISM and we have assumed a neutron star radius of 10 km.  Such bow shocks
have
recently been observed in optical (Cordes, Romani, \& Lundgren 1993) and X-ray
(Wang, Li, \& Begelman 1993) emission.  If $R_W$ shrinks inside the Bondi
radius
$$ r_B = 1.3 \times 10^{12} m v_{100}^{-2} \  {\rm cm} , \eqno(2)  $$
(where $m$ is the mass in solar units), then gravitational focusing will cause
the standoff distance
to collapse to the so-called Alfv\'en radius,
$$  r_A \sim 2\times 10^{10} B_{11}^{4/7} n_\infty^{-2/7} v_{100}^{6/7}  \ {\rm
cm} ,
\eqno(3) $$
where we have assumed that the density varies as $n_\infty (r / r_B)^{-3/2}$
inside
$r_B$.  Accretion onto the surface of the neutron star will then occur only if
the
corotation speed at $r_A$ is smaller than the local Keplerian speed,
corresponding to
$$P > 10^3 B_{11}^{6/7} n_\infty^{-3/7} v_{100}^{9/7}  \ {\rm s} ; \eqno(4) $$
otherwise, the gas will be prevented from accreting by the ``propeller
mechanism"
(Illarionov \& Sunyaev 1975).  Thus, only a very old (slowly rotating and
weakly magnetized)
neutron star is capable of accreting from the ISM.  Otherwise, $r_W$ or $r_A$
demarcates the boundary between the relatively dense circumstellar ISM and the
near-vacuum ``magnetospheric region".

The arguments given above apply only in the fluid limit.  Since $n_\infty
\ltorder 10$ cm$^{-3}$
typically (except during relatively rare passages through molecular cloud
complexes), the column
density of ISM spanning $r_B$ is usually less than $10^{14}$ cm$^{-2}$.  The
typical cross
section for inelastic collisions between neutrals is $\sim 10^{-16}$ cm$^2$ and
that between
neutrals and ions, due to charge exchange, can be an order of magnitude larger.
 Thus, if the
ISM is predominantly neutral at $r_B$ it will behave like a collisionless gas
rather than a fluid
(Begelman 1977), unless the timescale for photoionization is shorter than the
dynamical time.
Collisional ionization will be unimportant at the densities and particle
energies likely to be
present.  The density inside $r_B$ will then scale as $r^{-1/2}$, giving an
inward matter flux
$\propto r$. If there are no other UV sources, the photoionization time scale
will be very
sensitive to the surface temperature of the neutron star; temperatures
exceeding $\sim 10^5$ K
should be sufficient to ensure fluid-like behavior.  If ionization is weak, the
neutral component
of the ISM will be able to penetrate the pulsar wind and outer magnetosphere,
and may couple
to the magnetic field much closer to the neutron star, with implications for
the spindown rate and
wind properties of isolated pulsars.  However, hard photons emitted either in
the reverse shock
within the pulsar wind, or in the ISM bow shock, may also preionize the
neutrals, leading to a
self-consistent fluid-like structure.  There could also be circumstances where
the burst itself might
photoionize the ISM before (or immediately after) the blast wave hits it. The
importance of this
effect depends on how much UV comes out of the burst. To take a numerical
example (from
which one can do simple scaling) suppose that the burst emits a few times
$10^{35}$ ergs of
energy in the XUV. This corresponds to about $10^{46}$ photons which, at a
radius of order
$10^{13}$ cm, corresponds to a flux of  about $10^{19}$ cm$^{-2}$. Since the
photoionization
cross section is about $10^{-17}$ cm$^2$, most neutral atoms would be ionized
by this passing
shell of photons, and would be swept up by the relativistic flow.

\bigskip

\centerline{\sechead 3.  RADIATIVE PROPERTIES OF BLAST WAVES}

\medskip

Suppose an amount of energy $E_0 \sim 10^{39}E_{39}$ ergs is released in a
medium of number density $n(r)$ cm$^{-3}$, where $r$ is the distance from the
source
of the energy. This release may be assumed to be impulsive if it occurs over a
time
shorter than typical dynamical time scales in the subsequent flow.  The initial
energy produces
a highly relativistic fluid, with Lorentz factor $\eta$, if the mass $M_0$
initially released along
with the energy satisfies $E_0/M_0c^2  \equiv \eta \gg 1$.  After an amount
$\eta^{-1}M_0$
of external mass has been swept  up a blast wave forms ahead of the ejecta,
which starts to
decelerate.  In this decelerating regime, if radiation were inefficient the
bulk Lorentz factor of
the blast wave, after having reached the value $\Gamma\simeq \eta$, would vary
with radius
according to
$$\Gamma \sim \left({ 3 E_0 \over 4\pi m_p c^2 n r^3 }\right)^{1/2} . \eqno(5)
$$
The blast wave, however, may radiate away enough of its energy in a
sufficiently short
time scale to be of interest for explaining GRBs.  The most promising radiation
mechanism is
nonthermal synchrotron radiation by relativistic electrons accelerated at the
shock front which
propagates into the ISM, or present in the reverse shock which slows the ejecta
(M\'esz\'aros,
Laguna, \& Rees 1993).  The former can occur if the magnetic energy density is
amplified
behind the shock front (due to turbulent shear, etc.) to a significant fraction
($\lambda$) of
equipartition with respect to the shocked ambient gas.  In the latter case,
magnetic domination
is virtually guaranteed by the nature of the flow.  If the synchrotron
radiative efficiency
approaches one, the synchrotron-self-Compton losses also become important.
However, for
Galactic bursts, synchrotron is probably responsible
for most of the photons below 100 MeV, and detectable fluences can be obtained
even with
efficiencies as small as $10^{-3}$.

In the comoving frame, the magnetic field is given by
$$B' \sim 0.3 \lambda^{1/2} n^{1/2} \Gamma \ {\rm G} . \eqno(6) $$
\def\vm{\varepsilon_{MeV}}
To produce synchrotron photons of observed (Doppler-boosted) energy $\vm$ MeV
requires that electrons be accelerated to random Lorentz factors (in the fluid
frame) $\gamma$ such that
$$ \gamma\Gamma \sim 2.6\times 10^7
(\lambda n)^{-1/4} \vm^{1/2}  \ .  \eqno(7) $$
This is several orders of magnitude higher than the mean Lorentz factor per
electron,
$\sim (m_p/m_e) \Gamma$, which would apply if energy were shared equally among
all
particles behind the shock.  However, such an unequal distribution of energies
is
expected from models of Fermi acceleration behind strong shocks, which predict
that
a significant fraction of the shock energy can be pumped into the upper end of
the
relativistic electron energy distribution (Ellison \& Reynolds 1991).  The
synchrotron radiative efficiency of electrons accelerated in the blast wave, as
a function of $r$
and $\gamma$, is given by $\epsilon_{\rm rad} \sim \min [1, t'_{\rm exp}/
t'_{\rm syn}]$, where
$t'_{\rm exp} \sim r / c\Gamma$ is the expansion timescale of the blast wave
and
$t'_{syn} \sim 4\pi m_e c / \sigma_T (B')^2 \gamma $ is the synchrotron cooling
time,
both measured in the comoving frame.  Substituting from equations (6) and (7),
we have
$$  {t'_{\rm exp}\over  t'_{\rm syn} } \sim  1.5 \times 10^{-13}  (\lambda
n)^{3/4} r \vm^{1/2} \ .
\eqno(8) $$
Thus, for blast wave radii $\gtorder 10^{13}$ cm and typical interstellar
conditions $n \sim 1$
cm$^{-3}$, the emission of synchrotron radiation at energies above 1 MeV can be
highly
efficient.

The radiative efficiency of the blast wave at a given photon energy is
controlled by the
particle acceleration process, many details of which are uncertain.  The
maximum energy reached
by electrons is probably dictated by a balance between the acceleration and
cooling time
scales.  If we assume that shock acceleration to a Lorentz factor $\gamma$
requires $100
\zeta_2$ gyro-orbital times, then the maximum synchrotron photon energy coming
from the blast
wave is given by
$$ \varepsilon_{\rm max} \sim 0.4 \zeta^{-1}_2 \Gamma \ {\rm MeV} . \eqno(9) $$
Note that $\varepsilon_{\rm max}$ is independent of assumptions about the
magnetic field strength,
but does depend on the highly uncertain shock acceleration rate through
$\zeta$.  The overall
radiative losses from the blast wave are also affected by the fraction of shock
energy that goes
into relativistic electrons, as a function of radius.  If this is a fixed
fraction of the energy
dissipated in the shock, say, $dE/dr \sim - \mu E/r$, then the total energy in
the blast wave will
decrease as $E(r) \sim E_0 (r/r_0)^{-\mu}$, where $r_0$ is the radius at which
the shock starts
decelerating.  The bulk Lorentz factor is then given by eq.~(5) with $E(r)$
replacing $E_0$.
Photons of energy $\varepsilon$ will come predominantly
from inside the radius at which
$\varepsilon \sim \varepsilon_{\rm max}$, i.e., where the blast wave has slowed
to $\Gamma \sim
2.5 \zeta_2 \vm $.  In the limit $\mu \ll 1$, this radius is given
by $r_{\rm max} \sim 3\times
10^{13} (E_{39}/ n\zeta_2^2 \vm^2)^{1/3} $ cm, corresponding to a maximum burst
duration
of $\Delta t_{\rm max} \sim r/c\Gamma^2 \sim  160 (E_{39} / n)^{1/3} (\zeta_2
\vm)^{-8/3}$ s.
This estimate suggests that the maximum burst duration might be anticorrelated
with
energy of the observing band.  Note, however, that the extreme sensitivity to
$\zeta$ makes it
difficult to extract useful numerical estimates from this formula.

If the blast wave begins to decelerate at radii much smaller than $r_{max}$, a
significant flux
at energies $\sim \varepsilon$ could emerge on shorter timescales.  A rough
estimate for a
minimum timescale would correspond to the radius at which the synchrotron
radiative efficiency
(eq.~[8]) first approaches unity, $r_{\rm rad} \sim 6.7\times 10^{12} (\lambda
n)^{-3/4} \vm^{-
1/2}$ cm.  The Lorentz factor of the blast wave at this radius
(for $\mu \ll 1$) is
$\Gamma_{\rm rad} \sim 23 (E_{39}/ n )^{1/2} (\lambda n)^{9/8} \vm^{3/4}$ and
the characteristic
timescale is $\Delta t_{\rm rad} \sim 0.4 E_{39}^{-1} \lambda^{-3} n^{-2}
\vm^{-2}$ s.  A
necessary condition for the blast wave to radiate efficiently  at energy
$\varepsilon$ is that
$r_{\rm rad} < r_{\rm max}$, which is equivalent to the condition
$$n > 0.03 E_{39}^{-4/5} \lambda^{-9/5} \zeta^{8/5} \vm^{2/5}  \ {\rm cm}^{-3}
\ .
\eqno(10) $$
While the numerical values of the parameters in eq.~(10) are very uncertain,
the
condition suggests a correlation between burst efficiency (and therefore
detectability) and the density of the ambient ISM.

\bigskip

\centerline{\sechead 4. INFERENCES FROM BURST STATISTICS}

\medskip

If bursts repeat on a timescale of order $t_r$ years, then the local population
of
bursters comprises of order $10^3 t_r $ neutron stars.  Given a Galactic pulsar
birthrate of $\sim 10^{-11}$ pc$^{-2}$ yr$^{-1}$ (Narayan and Ostriker 1990),
the mean age of a bursting neutron star is $t_{\rm burst} \sim 10^7
R_{\rm kpc}^3
(t_r / f)$ yr $\equiv 10^{10} t_{10}$ yr, where $R_{\rm kpc}$ is the mean
distance to
bursts in kpc and $f$ is the fraction of the time during which the deposition
of burst
energy in the ISM would lead to a detectable burst.  Since the dipole spindown
time of
a pulsar is $\sim 10^9 P^2 B_{11}^{- 2}$ yr, the typical spin period of neutron
stars
responsible for the local bursts would be $\sim 3 B_{11} t_{10} $ s.  If
$t_{10} <
3 (v_{100} / B_{11})^{1/2} n_\infty^{-1/2}$, these pulsars would still be
producing
wind-driven bow shocks in the ISM, and would not be accreting interstellar gas.

The contact discontinuity between the shocked pulsar wind and the ISM would be
located
at  $r_W \sim 10^{13} B_{11}^{-1} t_{10}^{-1} v_{100}^{-1} n_\infty^{-1/2}$ cm.
 This
number is smaller than $r_{\rm max}$ for 1 MeV photons provided that $B_{11}
t_{10} n^{1/2}
> 0.3$, suggesting that detectable bursts from blast waves would come primarily
from a
relatively old population of pulsars, $t_{\rm burst} \gtorder 10^{9}$ yr,
and/or from neutron stars
passing through the denser regions of the ISM.  In either case, we estimate
$t_r / f \gtorder
100$.   Note that, in the simplest interpretation, $f$ would be the volume
filling factor of
ISM with high enough density to make the blast wave readily detectable.

\bigskip

\centerline{\sechead 5. DISCUSSION}

\medskip

Given recent renewed speculation about the distances of GRBs, we have extended
previous ideas
about plausible radiation mechanisms for Galactic GRBs, pointing out that
relativistic blast waves
driven into the ISM by magnetospheric disturbances around neutron stars can
yield bursts of
gamma-rays with roughly the observed range of timescales and fluences.  Our
extremely simple
conjectures about the radiative properties of synchrotron-emitting blast waves
do not reveal the
expected spectral properties of such bursts, but they do suggest a plausible
correlation between
the radiative efficiency at MeV energies and the density of the ambient medium.

The question of what might trigger gamma-ray bursts in this picture is
unresolved.
A model invoking neutron starquakes or other impulsive events that violently
disturb the
magnetosphere seems attractive on energetic grounds (Blaes et al.~1989).  The
energy in the
dipole magnetic field is $\sim 10^{39} B_{11}^2$ erg, which is enough for
individual bursts
but would require replenishment to explain frequent bursts.  One possibility is
the existence of a stronger non-dipole field close to the neutron star surface
(Ruderman 1993) which could transfer energy to the more distant magnetosphere.
Rotational
and, of course, gravitational energies would be adequate to power numerous
bursts per
neutron star.  Since we argue that the neutron stars responsible for GRBs are
probably not
accreting, triggers due to gas falling on the neutron star surface do not seem
as likely, although
starquakes triggered by asteroid or comet impacts are possible (Harwit \&
Salpeter 1973).

Whatever trigger mechanism leads to violently shaking of a neutron star
magnetosphere, we
argue that a strong gamma-ray burst can be generated by interaction of the
expanding energy
flow (whatever its form) with the ISM. This does not exclude a burst of
gamma-ray emission also
coming from the magnetosphere itself, but the latter is liable to be highly
variable, shorter and
probably less  efficient.  In the light of the evidence for two classes of
classical GRBs
(Kouveliotou et al.~1993; Lamb et al.~1993), it would seem plausible to
attribute the short
bursts to the latter mechanism and the long ones ($\Delta t \gtorder 2$ s) to
the blast wave. The
radiative efficiency of the blast wave depends on the density of the ISM (the
structure of which
may introduce longer timescale variability), so the variable component may be
present in all
bursts but the smooth component would be dominant (and overwhelm the former)
for bursts
occurring in regions where the ISM has reasonably large density (e.g., clouds,
not necessarily
molecular).

Old pulsars are expected to have a smooth distribution in the Galaxy,
constituting a halo
population or a disc population with a large scale height.  If the bursts came
from distances
$\gtorder 10$ kpc, one would expect a strong systematic concentration towards
the Galactic
Centre; on the other hand, if the burst distances were $\sim 1$ kpc or less (as
has been favoured
by most previous theoretical treatments involving Galactic models) and their
distribution directly
traced that of the old neutron stars, the non-uniformities revealed by $V/V_m$
would be
perplexing.  A $\log N -\log S$ slope flatter than 3/2 at low fluences can be
easily understood
in terms of a dropoff in the number of sources beyond a local density excess
associated with our
immediate neighbourhood. However, it seems a bit of a coincidence that the
anisotropy is rather
small relative to the deficit from Euclidean counts --- this implies that we
are relatively near the
centre of a kpc-scale region where the mean ISM density is higher than on the
outside.

This model, based on a local burst population made more conspicuous by a denser
gaseous
environment, would predict that the spatial distribution of smooth-burst
sources should be
modulated by the highly irregular and structured distribution of the ISM. The
spiral-arm effects
discussed by Quashnock \& Lamb (1993) are a natural consequence of our
proposal.  Our
explanation for these effects is more plausible than an alternative explanation
attributing all bursts
to neutron stars just a few million years old which still remember the spiral
arm they came from,
since the latter would require a much higher repetition rate.

\bigskip
We acknowledge partial support from NSF grant AST91-20599 and NASA grant
NAG5-2026 (MCB), from NASA NAGW-1522 (PM), and from the Royal Society (MJR).

\bigskip

\centerline{\sechead REFERENCES}

\def\ref{ \parindent=0pt
\hangindent=20pt
\hangafter=1 \smallskip}

\medskip

\ref Begelman, M.~C. 1977, MNRAS, 181, 347
\ref Blaes, O., Blandford, R., Goldreich, P., \& Madau, P. 1989, ApJ, 343, 839
\ref Cordes, J.~M., Romani, R.~W., \& Lundgren, S.~C. 1993, Nature, 362, 133
\ref Ellison, D.~C., \& Reynolds, S.~P. 1991, ApJ, 382, 242
\ref Fishman, G., et al. 1993, ApJS, in press
\ref Harwit, M., \& Salpeter, E. E.  1973, ApJ, 186, L37
\ref Illarionov, A., and Sunyaev, R. 1975, A\&A, 39, 185
\ref Kouveliotou, C., et al. 1993, ApJ, 413, L101
\ref Lamb, D.~Q., Graziani, C., \& Smith, I. 1993, ApJ, 413, L11
\ref Meegan, C., et al. 1992, Nature, 355, 143
\ref M\'esz\'aros, P., Laguna, P., \& Rees, M.~J. 1993, ApJ, in press
\ref Narayan, R., and Ostriker, J.~P. 1990, ApJ, 352, 222
\ref Quashnock, J.~M., \& Lamb, D.~Q. 1993, MNRAS, in press

\ref Ramaty, R., Bonazzola, S., Cline, T.~L., Kazanas, D. \& M\'esz\'aros, P.
1981,
    Nature, 287, 122
\ref Ruderman, M. 1993, ApJ, in press
\ref Shull, J.~M., Fesen, R.~A., \& Saken, J.~M. 1989, ApJ, 346, 860
\ref Wang, Q.~D., Li, Z., \& Begelman, M.~C. 1993, Nature, 364, 127

\end